\documentclass[12pt]{article}

\usepackage{amsmath,amssymb}

\newcommand{\be}{\begin{equation}}
\newcommand{\ba}{\begin{eqnarray}}
\newcommand{\ea}{\end{eqnarray}}

\def\d{\delta}

\def\f{\phi}

\def\l{\lambda}

\def\p{\pi}

\def\F{\Phi}

\def\ca{{\cal A}}

\def\cd{{\cal D}}

\def\cf{{\cal F}}
\def\cg{{\cal G}}
\def\ch{{\cal H}}

\def\cl{{\cal L}}
\def\cm{{\cal M}}

\def\cp{{\cal P}}

\newcommand{\pa}{\partial}

\newtheorem{thm}{Theorem}[subsection]

\newtheorem{proposition}[thm]{Proposition}

\newcommand{\bbN}{{\Bbb N}}
\newcommand{\C}{{\Bbb C}}

\newcommand{\bbC}{{\Bbb C}}
\newcommand{\bbH}{{\Bbb H}}

\newcommand{\bbR}{{\Bbb R}}

\newcommand{\bbT}{{\Bbb T}}

\newcommand{\One}{{\boldmath 1}}

\newcommand{\Hom}{\operatorname{Hom}}

\newcommand{\id}{\operatorname{id}}

\fontfamily{yfrak}

\begin{document}

\vskip 15mm

\begin{center}

{\Large\bfseries Spectral triples of holonomy loops 
}

\vskip 4ex

Johannes \textsc{Aastrup}$\,^{1}$ \&
Jesper M\o ller \textsc{Grimstrup}\,$^{2,3}$ 

\vskip 3ex  

$^{1}\,$\textit{Institut f\"ur Mathematik, Universit\"at Hannover \\
  Welfengarten 1, D-30167 Hannover, Germany}
\\
e-mail:  
\texttt{aastrup@math.uni-hannover.de}
\\[3ex]
$^{2}\,$\textit{NORDITA \\Blegdamsvej 17, DK-2100 Copenhagen, Denmark}
\\
e-mail: \texttt{grimstrup@nordita.dk}
\\[3ex]
$^{3}\,$\textit{Science Institute, University of Iceland \\ Dunhaga 3, IS-107 Reykjavik, Iceland}
\end{center}

\vskip 5ex

\begin{abstract}

  The machinery of noncommutative geometry is applied to a space of
  connections. A noncommutative function algebra of loops
  closely related to holonomy loops is investigated. The space of connections
  is identified as a projective limit of Lie-groups composed of copies of the
  gauge group.  A spectral triple over the space of connections is obtained by
  factoring out the diffeomorphism group. The triple consist of
  equivalence classes of loops acting on a hilbert space of sections
  in an infinite
  dimensional Clifford bundle. We find that the Dirac operator acting on this
  hilbert space does not
  fully comply with the axioms of a spectral triple.

\end{abstract}

\newpage

\tableofcontents
\newpage
\section{Introduction}

The story of noncommutative geometry starts with the idea that instead of
studying spaces one studies  algebras of functions on the spaces. A concrete
result supporting this idea is the Gel'fand-Naimark theorem \cite{gelfand} that states that the world of
locally compact Hausdorff spaces is, by taking the corresponding algebra of continuous
complex valued functions vanishing at infinity, the same as
the world of commutative $C^*$-algebras. Hence noncommutative $C^*$-algebras
can be considered as noncommutative locally compact Hausdorff spaces.

The crucial leap from noncommutative topology to geometry was done by Alain
Connes \cite{ConnesBook}. The key observation is that the Dirac operator on a Riemannian
manifold gives full information about the metric. This idea provides the
definition of a noncommutative geometry, i.e. a spectral triple, by
abstractizing a Dirac operator as an operator acting on the same hilbert space
as the (non)commutative algebra; satisfying a list of axioms generalizing
interaction rules of smooth functions with the Dirac operator.
 
Prime examples of noncommutative geometries are given by quotient spaces. A 
conceptually simple case is the set of two points identified. The classical
way of identification would be to consider just one point. The noncommutative
quotient is to consider two by two matrices. So we regard the two
sub-algebras 
\ba
\left(
\begin{array}{cc}
\bbC & 0 \\
0 & 0
\end{array}
\right)\;,
\left(
\begin{array}{cc}
0 & 0 \\
0 & \bbC
\end{array}
\right)\nonumber
\ea
as the function algebras over the two points. These algebras are then
identified through the partial isometries 
\ba
\left(
\begin{array}{cc}
0 & 1 \\
0 & 0
\end{array}
\right)\;,
\left(
\begin{array}{cc}
0 & 0 \\
1 & 0
\end{array}
\right)\nonumber\;,
\ea
which not only identify the points but also belong to the
algebra. Represented on $\ch=\bbC\oplus\bbC$ the algebra of two by two
matrices interacts with a Dirac operator given by
\ba
\displaystyle{\not}\cd=\left(
\begin{array}{cc}
0 & a \\
-a & 0
\end{array}
\right)\nonumber\;,\quad a\in\bbR\;.
\ea

This noncommutative geometry, when combined with the commutative algebra of smooth functions on a
manifold, is related to the Higgs effect in the Connes-Lott model
\cite{Connes:1990qp} and to the Higgs effect in
Connes' full formulation of the Standard Model \cite{Connes:1996gi}. The crucial point is
that exactly the noncommutativity of the algebra generates the entire bosonic
sector, including the Higgs scalar,
through fluctuations around the Dirac operator. The action of the standard model coupled
to gravity comes out \cite{Chamseddine:1991qh,Chamseddine:1996rw}
\[
\langle\xi |
\tilde{\displaystyle{\not}\cd} |\xi \rangle + \mbox{Trace} \;\varphi\left( \frac{\tilde{\displaystyle{\not}\cd}}{\Lambda}\right)\;,
\]
where $\tilde{\displaystyle{\not}\cd}$ is the fluctuated Dirac operator, $\xi$
a hilbert state and $\varphi$ a suitable cutoff function selecting eigenvalues
below the cutoff $\Lambda$. 

Unfortunately, this beautiful unification of the standard model with general relativity is
completely classical. No clear notion of quantization exists within the
framework of noncommutative geometry.\\

The aim of this paper is to explore new ideas on the unification of noncommutative geometry with
the principles of quantum field theory. 

Quantum field theory deals with spaces of field configurations. The central object is the path integral 
\[
\int \cd\F\exp{\left(\frac{\rm{i}}{\hbar}S[\F]\right)}\;,
\]
where $\F$ denotes the field content of the theory described by the
(symmetries of the) classical action
$S[\F]$. $\cd\F$ is a formal measure on the space of field
configurations. Therefore, rather than dealing with manifolds or 
algebras of functions hereon, quantum field theory lives on the much larger spaces of field
configurations. 
We now suggest the following: If Connes' formulation of the standard model and quantum field theory are to be linked, and if the
principles of noncommutative geometry are fundamental (which we believe they are), then one should apply the machinery of noncommutative geometry
to some space of field configurations. Further, since Connes' formulation of
the standard model is in principle a gravitational theory (pure geometry) we suggest that
the correct implementation of quantum theory must involve quantum
gravity. Thus, we suggest to study a functional space related to
general relativity. 

The aim is to find a suitable configuration space on which a generalized
Dirac operator exist. A function algebra hereon may very well be naturally
noncommutative (classically). The hope is that the Dirac operator will
generate a kind of quantization of the underlying space.

For the space of field configurations we use ideas from loop quantum gravity \cite{Ashtekar:2004eh}. Here the space is
the space of certain connections modulo gauge equivalence. The function algebra is
generated by traced holonomies of connections along loops, i.e. all physical observables can
be expressed by these. This gives a commutative
algebra. However, the lesson taught by noncommutative geometry is that
the noncommutativity of the algebra provides essential structure. The idea is
therefore to keep the noncommutativity by taking holonomies without tracing
them; a loop $L$ maps connections into group elements of $G$
\ba
L:\nabla\rightarrow Hol(L,\nabla)\in G\;,
\label{first}
\ea
where $Hol(L,\nabla)$ is the holonomy along $L$, and $G$ is the gauge
group which we, for now, assume to be compact. Loops functions like (\ref{first})
 correspond to an underlying space of gauge connections which includes also gauge
 equivalent connections.
 This will also resemble Connes'
construction of standard model, since we get an algebra of matrix valued functions over a
configuration space just as Connes' matrix valued functions over a manifold.

Furthermore, in loop quantum gravity a fibration of the space-time
manifold into global space and time directions is considered. This is done in order to apply
a canonical quantization scheme. In the present case the aim is to construct a
spectral triple over a functional space of connections. For this purpose such a
fibration is not needed and we therefore consider the whole
manifold. Thus, the connections considered are space-time connections.

The central achievement of loop quantum gravity is its ability to obtain a
separable hilbert space of loop functions via diffeomorphism
invariance\footnote{In fact, diffeomorphism invariance alone does not give a
  separable hilbert space. Instead one has to use a generalized notion of
  diffeomorphisms, see \cite{Fairbairn:2004qe}.} (see \cite{Fairbairn:2004qe} and references therein). It is possible to extend these results to the case of a
noncommutative algebra; we represent certain equivalent classes of noncommutative loop
operators on a diffeomorphism invariant, separable hilbert space. Further,
the Dirac operator we construct on the holonomy algebra is diffeomorphism invariant and hence also
descends to the diffeomorphism invariant hilbert space. This is important
since the Dirac operator stores the full physical information. \\

Let us finally add a note on noncommutativity and quantum theory. Clearly, the
noncommutativity suggested is classical: It is simply related to the
non-Abelian structure of the group $G$ and therefore carries no quantum
aspect. On the other hand, the Dirac operator which we construct will resemble
a global functional derivation. As a Dirac operator it carries spectral
information of the underlying space -- the space of connections -- and will
enable integration theory. In this sense, the quantum aspect enters through
the constructed Dirac operator.\\

\noindent{\bf Outline of the paper}\\

The algebra of (untraced) holonomy transformations, which is a central object in
this paper, is introduced as the hoop group $\ch\cg$ in section
\ref{sec-hoopgroup}. Since a smooth connection in a $G$-bundle maps a loop $L\in\ch\cg$
into $G$ homomorphically via the holonomy transform we define in section \ref{Rep} the space $\bar{\ca}$ of
generalized connections as the set of homomorphisms
\[
\bar{\ca}= Hom(\ch\cg,G)\;.
\] 
This is the functional space on which we wish to do geometry.
Conversely, since the hoop group acts on $\bar{\ca}$ simply by 
\[
H_L(\nabla)= \nabla(L)\;,
\]
we interpret $\ch\cg$ as a noncommutative function algebra on $\bar{\ca}$. The
key technical tool for dealing with the space $\bar{\ca}$ is described in
section \ref{sec-projectivelimit}. Referring to \cite{Marolf:1994cj} we identify $\bar{\ca}$ as a projective limit
over the representations of finite subgroups of the hoop group. This enables
us to work with only finitely many loops at a time. The space $\bar{\ca}$ seen
from finitely many loops looks like
\[
G^n= \underbrace{G\times\ldots\times G}_{\mbox{\small $n$ times}}\;,
\] 
where $n$ is related to the number
of loops. Thus, since $G$ is a Lie-group, we are at this level dealing with just an ordinary manifold and
we can therfore write down Dirac operators from classical geometry. 

A concrete realization of this technique/idea is worked out in  section
\ref{hilbert-section}. Since we are sitting in a projective system we are not entirely free
to choose our Dirac operator; it has to fit with different choices of finitely
many loops. In fact, problems arise from loops with common line segments. We
remedy this defect by technically excluding such combinations of loops. Also,
for technical reasons, we choose the classical Euler-Dirac instead of the real
Dirac operator.

In doing this the link to connections becomes unclear. This is clarified in
section \ref{section-connection}, where we show that the connection are still contained in the
spectrum of the modified algebra.

A key issue in the construction presented is the implementation of
diffeomorphism invariance; the concern of section \ref{section-diffeomorphism}. Using once more
ideas from loop quantum gravity we construct diffeomorphism
invariant states and a diffeomorphism invariant algebra of loop
operators. Finally we are concerned with the question whether the obtained,
diffeomorphism invariant triple is spectral in the sense of Connes. It turns
out not to be the case since the eigenvalues of the Dirac operator has infinite
multiplicity. In particular, this is linked to the kernel of the Euler-Dirac
operator on $G$ which has dimension larger than one. Although we are at
present unable to solve the problem we suggest some possible solutions.

We provide a final discussion and outlook in section \ref{sec-dis} and leave some
extra material for the appendices.

\section{The hoop group}
\label{sec-hoopgroup}

The starting point is a manifold $M$. Let us for simplicity assume
that $M$ is topological trivial. On this manifold we
consider first the set $\cp$ of
piecewise analytic paths
\[
\cp :=\big\{ P(t)|P:[0,1]\rightarrow M  \big\}\;,
\]
where paths which differ only by a reparameterization
are identified. If two paths $P_1,P_2\in\cp$ have coinciding end and start
points, $P_1(1)=P_2(0)$, we define their product
\[
P_1\circ P_2( t) = \left\{
\begin{array}{ll}
P_1(2 t)  &  t\in[0,\tfrac{1}{2}]\vspace{2mm}  \\
P_2(2 t-1)&  t\in[\tfrac{1}{2},1]
\end{array}\right.\;.
\]
In case $P_1(1)\not=P_2(0)$ we set their product is zero. There is a natural involution on $\cp$
\[
P^\ast(t)=P(1-t)\qquad\forall t\;,
\]
since
\[
 (P^\ast)^\ast =P\;,\qquad (P_1\circ P_2)^\ast= P_2^{\ast}\circ P_1^\ast\;.
\]
Choose an arbitrary basepoint $x^o\in M$. We call a path which starts and ends at $x^o$ a based
loop. Further, by a simple loop we understand a based loop for which
\[
L(t)=x^o \Leftrightarrow t\in \{0,1 \}\;.
\]
 The set of
based loops is called loop space and is denoted $\cl_{x^o}$.
An equivalence relation on loop space is generated by identifying
loops which differ by a simple
retracing along a path   
\[
L_1 \sim L_2 \Leftrightarrow 
\left\{
\begin{array}{l}
L_1 = P_1 \circ P_2 \circ P_2^\ast \circ P_3\\
L_2 = P_1\circ P_3
\end{array}\right.\;,
\]
where $L_i\in\cl_{x^o}$, and $P_i\in\cp$. An equivalence class $[L]$ is called a {\it hoop} 
\cite{Ashtekar:1993wf}. The set of hoops is called the {\it hoop group},
denoted 
\[
\ch\cg=\cl_{x^o}/\sim\;,
\] 
since the involution on
$\ch\cg$ gives an inverse element
\[
[L]\cdot[L]^\ast =[L_{\id}]\;,
\]
where $L_{\id}$ is the trivial loop $$L_{\id}(t)= x^o\quad\forall t\in[0,1]\;.$$ To ease the
notation we will denote a hoop $[L]$ simply by an representative $L$ of the
equivalence class. Furthermore, for literary reasons we often call $[L]$ a loop.

We emphasize that since $M$ has no metric any notion of distance between and
length of loops
and hoops
is meaningless.

\section{Hoop group representations}
\label{Rep}

Consider the space of homomorphisms
\[
\bar{\ca}= \mbox{Hom}(\ch\cg,G)\;,
\]
from the hoop group into a matrix representation of a compact Lie group $G$ (we denote both the
group and its representation by $G$. The group $G$ is assumed to have a
both left and right invariant metric). That is, for $\nabla\in\bar{\ca}$ we
have
\[
\nabla(L_1)\cdot\nabla(L_2)=\nabla(L_1\circ L_2)\qquad\forall\; L_1,L_2\in\ch\cg\;.
\]
 If we
denote by $\ca$ the space of smooth connections in a bundle with structure group $G$,
then a connection $\nabla\in\ca$ clearly gives such a homomorphisms via
\ba
\nabla:L\rightarrow Hol(L,\nabla)\;,
\label{map-nabla}
\ea
where $Hol(L,\nabla)$ is the holonomy of the connection around the loop
$L$. Let us recall that the holonomy is the parallel
transport of the connection along a path $P$
\[
Hol(P,\nabla)=\cp \exp\left({\rm{i}}\int_P \nabla\right) \;,
\]
where $\cp$ is the path ordering symbol. The parallel transport along a closed
loop is a non-local, gauge covariant object and the trace
hereof, the Wilson loop, is gauge invariant. 
From (\ref{map-nabla}) we conclude that
\[
\ca\subset\bar{\ca}\;.
\]
It is however important to realize that
$\bar{\ca}$ is much larger\footnote{In fact, $\ca$ has, with respect to the
  Ashtekar-Lewandowski measure, zero measure in $\bar{\ca}$ (modulo gauge
  transformations). } \cite{Marolf:1994cj}.

The space $\bar{\ca}$ is the general space of field configurations on which we
wish to obtain a geometrical structure. We therefore consider an
algebra of functions over $\bar{\ca}$. To this end we first notice that
a hoop $L\in\ch\cg$ gives rise to a function $H_L$ on $\bar{\ca}$ into $G$ via
\ba
H_L(\nabla)=\nabla(L)\;,
\label{generators}
\ea
where $\nabla\in\bar{\ca}$. Notice that 
\ba
H_{L_1}\cdot H_{L_2}=H_{L_1\circ L_2}\;,\quad
(H_{L})^\ast = H_{L^{-1}}\;,\quad
H_{L_{\id}}= \One\;,\nonumber
\ea
where $L_i\in\ch\cg$. The set of complex linear combinations of all functions
$H_L$ is a $\star$-algebra. The norm of a general linear combination
\[
a_1 H_{L_1}+ \ldots + a_n H_{L_n}
\]
is defined by
\[
\| a_1 H_{L_1}+ \ldots + a_n H_{L_n} \| = \sup_{\nabla\in\bar{\ca}}\left(\|a_1
H_{L_1}(\nabla)  + \ldots + a_n H_{L_n}(\nabla)\|\right)\;,
\]
where $\| \cdot\|$ on the rhs is the matrix norm. Notice that 
\[
\| H_L \| =1\quad \forall L\in\ch\cg
\]
if the group $G$ is orthonormal or unitary. The closure in this norm of the algebra generated by functions $H_L$ is a $C^\star$-algebra.
Let us denote it $C^\ast(\cl_{x^o})$. For now, this is the noncommutative function algebra
over $\bar{\ca}$ which we wish to imbed in a spectral triple. However, as we
will explain in the next section, we need to change the algebra slightly to be
able to construct a Dirac operator.

\section{The space $\bar{\ca}$ as a projective limit }
\label{sec-projectivelimit}

The space $\bar{\ca}$ was analyzed in \cite{Marolf:1994cj} in a somewhat different
context\footnote{The authors of \cite{Marolf:1994cj} considered the
  smaller space $\bar{\ca}/Ad$ of smooth connections modulo local gauge
  transformations. This otherwise important difference is not essential for
  the issues regarding the projective limit.}. Here, the authors identify $\bar{\ca}$ with a projective
limit (see appendix \ref{proj} for details on projective and inductive limits)
\[
\lim_{\leftarrow} Hom(F,G)\;,\qquad F\in\cf\;,
\] 
where $\cf$ is the set of all strongly independent, finitely
generated subgroups of $\ch\cg$ (strongly independent in the sense of \cite{Ashtekar:1993wf}).
Let $L_1\ldots L_{n(F)}$ be the strongly independent generators of
$F\in\cf$. We 
then identify 
$$Hom(F,G)\simeq G^{n(F)}$$ since we just map $\f\in Hom(F,G)$
into \cite{Ashtekar:1993wf} 
\ba
(\f(L_1),\ldots, \f(L_{n(F)}))\in G^{n(F)}\;.
\label{identification}
\ea
This identification is of great advantage: Since $G^{n(F)}$ is a Lie group it is now
straight forward to construct a spectral triple by choosing a metric on $G$
and then using the Euler-Dirac\footnote{see appendix \ref{diracappendix}.} or the
Dirac operator (since $G^{n(F)}$ is a Lie group it is parallelizable and hence
possesses a spin structure). Once a geometrical construction on $G^{n(F)}$ is
obtained we extend this to all of $\bar{\ca}$ by taking the projective limit of
the algebra and the inductive limit of the relevant hilbert space. 

Thus, it is tempting to consider the hilbert
space 
\[
L^2(G^{n(F)},M_k(\bbC)\otimes S)\;,
\] 
where $S$ is the Clifford algebra or the spin bundle corresponding to either
the Euler-Dirac or the Dirac operator. $L^2$ is with respect to the Haar
measure on $G^{n(F)}$. The problem with this construction is
that the Euler-Dirac and the Dirac operators contains all the metric information on
the underlying space \cite{ConnesBook} and the structure maps defining the
projective limit are not metric. The problems can be traced back to the
definition of the generating hoops. Here, following \cite{Ashtekar:1993wf}, we
encounter overlapping hoops which leads to structure maps 
\[
P_{F_1,F_2} : Hom(F_1,G)\rightarrow Hom(F_2,G)
\]
of the form\footnote{here $Hom(F_1,G)$ and $Hom(F_2,G)$ are, as an example, identified with
  $G^1$ and $G^3$, respectively. A similar structure map with a $G^2$-subgroup is not possible due to the
  special construction of independent hoops.}
\ba
(g_1,g_2,g_3) \rightarrow g_1 g_2\;,
\label{structuremap}
\ea
where $F_1\subset F_2$ lie in $\cf$. The problem is that such maps do not have
a canonical isometric cross-section. 

The solution to this problem is to redefine our notion of generating
hoops. This, in turn, will affect the projective limit. Let us go
into details in the next section.

Before we do that we end this section by mentioning that the
identification (\ref{identification}) indirectly chooses an orientation of the
hoop. Basically, there are two possible identifications corresponding to
either $\varphi(L)$ or $\varphi(L^{-1})$. Therefore, we can identify
$Hom(F,G)$, where $F$ is a subgroup generated by a single hoop, with both $G$
and $G^{-1}$.

\section{Spectral triples over $G^n$ and the projective limit}
\label{hilbert-section}

Let $\cf_I$ be the set of finitely generated subgroups of $\ch\cg$ with the
property that they are generated by simple, non-selfintersecting loops that do
not have overlapping segments or points\footnote{In contract to
  \cite{Ashtekar:1993wf} we no-longer require loops to be piece-wise
  analytic. Nor does the manifold need a real analytic structure.}. The inclusion of groups $F_1\subset F_2$ gives
an inductive system on $\cf_I$ and therefore a projective structure on $\{
\Hom ( F,G)\}_{F\in \cf_I}$. Again, we can identify $\Hom (F,G)$ with
$G^{n(F)}$, where $n(F)$, as before, is the number of simple loops in a generating set of $F$. 

Since we are looking at subgroups with the property that no two loops have
overlapping segments the maps $P_{F_1,F_2}:G^{n(F_2)} \to G^{n(F_1)}$
induced by the inclusion $F_1 \subset F_2$ are just given by deleting some
coordinates or inverting some coordinates. This eliminates structure maps of
the form (\ref{structuremap}) and thus enables the following construction of a 
spectral triple.

\subsection{The hilbert space}

We first construct the hilbert space. We choose a left and right invariant
metric on $G$. We therefore also have a metric on $G^{n(F)}$ and hence we can
construct the Clifford bundle $Cl(TG^{n(F)})$. Due to the invariance of the
metric we get the result
\begin{proposition}
\label{prop}
There is  an embedding of hilbert spaces 
$$P_{F_1,F_2}^* :L^2(G^{n(F_1)},Cl(TG^{n(F_1)})) \to L^2(G^{n(F_2)},Cl(TG^{n(F_2)}))\;,$$
where the measure on $G^{n(F_i)}$ is the Haar measure. 
\end{proposition}

\textit{Proof.} We will need some notation. Let $e_1,\ldots , e_n$ be an
orthonormal basis in $T_{id}G$, the tangent space over the identity in
$G$. Due to the invariance property of the metric we get that $$D_g(id)
(e_1),\ldots , D_g(id)(e_n)$$ is an orthonormal basis in $T_gG$. Here
$D_g(id)$ denotes the differential of the map 
\[
m_g:G\to G\;,\quad m_g(g_1)= g g_1
\]
in the identity. We will also use the notation $e_1,\ldots, e_n$ to denote the corresponding global vector fields in $TG$, i.e. $e_k(g)=D_g(id)(e_k)$.

We will abbreviate $n(F_i)$ by $n_i$. We first consider the case where the projection $P_{F_1,F_2}$ is of the form 
\begin{equation}\label{strucaf1} P_{F_1,F_2} (g_1,\ldots ,g_{n_2})= (g_1,\ldots , g_{n_1})\;,
\end{equation}
and denote by 
$$e_1^1,\ldots , e_n^1,e_1^2, \ldots , e_n^2, \ldots ,e_1^{n_i}, \ldots , e_n^{n_i}$$  
the global vector fields on $G^{n_i}$, where $e_1^k , \ldots , e_n^k$ denote the global vector fields $e_1,\ldots , e_n$ in the $k$'th component of $TG^{n_i}$.  

Put $\bar{g}_{n_i}=(g_1, \ldots , g_{n_i})$. We clearly have that 
\ba
\langle e_l^k (\bar{g}_{n_1}),e_{l'}^{k'} (\bar{g}_{n_1})\rangle_{T_{\bar{g}_{n_1}}G^{n_1}}
=\langle e_l^k (\bar{g}_{n_2}),e_{l'}^{k'} (\bar{g}_{n_2})\rangle_{T_{\bar{g}_{n_2}}G^{n_2}}\;, 
\nonumber
\ea
where $k,k'\leq n_1$.

An element in $L^2(G^{n_1},Cl(TG^{n_1}))$ is a linear combination of elements of the form $fe$, where $e$ is a product of elements in 
$$e_1^1,\ldots , e_n^1,e_1^2, \ldots , e_n^2, \ldots ,e_1^{n_i}, \ldots , e_n^{n_1},$$
and $f \in L^2(G^{n_1})$. We define 
$$P_{F_1,F_2}^*(fe)=\tilde{f}e\;,$$
where 
\[
\tilde{f}(\bar{g}_{n_2})\equiv f(P_{F_1,F_2}(g_{n_2}))=f(\bar{g}_{n_1})\;.  
\]
This map preserves the inner product since
\begin{eqnarray*}
& \langle fe ,f'e'\rangle_{L^2(G^{n_1},Cl(TG^{n_1}))}\\
&= \int \bar{f}(\bar{g}_{n_1})f'(\bar{g}_{n_1}) 
\langle e,e'\rangle_{T_{(\bar{g}_{n_1})}G^{n_1}}\cdot 
d\mu_H(g_1)\cdots d\mu_H(g_{n_1}) 
\\
&= \int \bar{f}(\bar{g}_{n_1})f'(\bar{g}_{n_1}) 
\langle e,e'\rangle_{T_{(\bar{g}_{n_2})}G^{n_2}}\cdot 
d\mu_H(g_2)\cdots d\mu_H(g_{n_2}) \\
&=  \langle \tilde{f}e ,\tilde{f'}e'\rangle_{L^2(G^{n_2},Cl(TG^{n_2}))}\;,
\end{eqnarray*}
where we have used that $\int 1d\mu_H=1$.

To finish the construction we only need to consider a map of the form 
\begin{equation}
P_{F_1,F_2}(g)=g^{-1}
\label{strucaf2}\;,
\end{equation}
since any structure map is the composition of maps of the type (\ref{strucaf1}) and (\ref{strucaf2}).
However, the map
$$P_{F_1,F_2}^* :L^2(G,Cl(TG)) \to L^2(G,Cl(TG))\;,$$
defined by (with the notation from before)
\ba
P_{F_1,F_2}^* (fe)(g)=f(g^{-1})D_{P_{F_1,F_2}^{-1}}(e)
\nonumber
\ea
is, due to the left and right invariance of the metric, a map of hilbert
spaces. This completes the proof.\\

We can now construct the direct limit of these hilbert spaces (see appendix
\ref{proj} for a more detailed discussion on inductive limits). This is done in the following way: 
First define 
$$\ch_{alg}=\oplus_{F\in \cf_I}L^2 (G^{n(F)},Cl(TG^{n(F)}))/N\;, $$ 
where $N$ is the subspace generated by elements of the form 
$$(\ldots , v, \ldots , -P^*_{F_1,F_2}(v),\ldots )\;.$$ In other words, we
identify the vectors $v$ and $P^*_{F_1,F_2}(v)$.

The problem is now to define an inner product on $\ch_{alg}$. Decompose
$L^2(G,Cl(TG))$ into the subspace generated by the function $1$ and the
orthogonal complement. We will write this as 
$$L^2(G,Cl(TG))=\ch_1\oplus\ch_2\;,$$
where $\ch_1=\bbC$. Given a vector $v\in L^2(G^{n(F)},Cl(TG^{n(F)}))$ this can be uniquely
decomposed into vectors of the form 
\[
v_1\otimes\cdots\otimes v_{n(F)}\;,
\]
where each $v_i$ belong either to $\ch_1$ or $\ch_2$. It is therefore enough
to define the inner product of vectors of this type. Further, let $v_1\in
L^2(G^{n(F_1)},Cl(TG^{n(F_1)}))$ and $v_2\in
L^2(G^{n(F_2)},Cl(TG^{n(F_2)}))$ be vectors of this form. We will assume that
in the tensor decomposition of $v_1$ and $v_2$ only elements from
$\ch_2$ appear. We can assume this since else  $v_1$ and/or $v_2$ will be the image under
one of the $P$'s, and we can simply pull the vector back. We finally define the
inner product by
\ba
\langle v_1,v_2\rangle = \langle P_{F_1,F_3}^*
(v_1),P_{F_2,F_3}^*(v_2)\rangle_{L^2(G^{n(F_3)},Cl(TG^{n(F_3))})}
\label{inpr}
\ea  
if there exist a $F_3$ with $F_1,F_2 \subset F_3$ and zero else. The
completion of the $\ch_{alg}$ with respect to this inner product is the
inductive limit and will be denoted by $\ch'_{si}$ ({\bf si} $\sim$ segment independent).  

In equation (\ref{inpr}), since $v_1$ and $v_2$ are, per definition, decomposed into
tensor-powers in $\ch_2$, the inner product will be different from zero only when $F_1=F_2$.

We can give a more concrete description of $\ch'_{si}$ in terms of the
hilbert space $\ch_2$, namely
\ba 
\ch'_{si}=\C \oplus (\oplus_{l^1}\ch_2) \oplus
(\oplus_{l^2}\ch_2\otimes \ch_2)\oplus
\ldots\;,
\label{h2}
\ea
 where $l^k$ is the set of all products of $k$-nonintersecting simple loops,
and where $\oplus$ means orthogonal sum. The first $\C$ corresponds to the
trivial loop. For each simple loop we get a copy of $L^2(G,Cl(TG))$; however
the constant functions are identified in the inductive limit, and we hence only get a
copy of $\ch_2$ for each simple loop. This picture continues for products of
two simple loops and so on.

\subsection{The Euler-Dirac operator}

On each of the hilbert spaces $L^2(G^{n(F)},Cl(TG^{n(F)}))$ we have a
 canonical Euler-Dirac operator 
\ba
\displaystyle{\not} \cd (\xi)= \sum e_i\cdot \nabla_{e_i}(\xi)\;,
\label{EDi}
\ea
where $\{e_i\}$ are global, orthonormal sections in the tangent bundle of
$G^{n(F)}$ and $\nabla$ is the Levi-Civita connection.
 It is clear that this Euler-Dirac operator commutes with the structure maps
 $P_{F_1,F_2}^*$ not involving inversions. According to \cite{spingeometri}
 $\displaystyle{\not}\cd$ can, under the identification of $Cl(TM)$ with
 $\wedge^* (T^*M)$ (differential forms), be identified with $d+d^*$.
  The exterior derivative $d$ is invariant under all diffeomorphisms, and since
  $d^*$ only additionally depends on the metric and the metric on $G$ is
  invariant under inversions, the Euler-Dirac operator also commutes with
  structure maps involving inversions. Therefore get an Euler-Dirac operator $\displaystyle{\not} \cd $ on $\ch'_{si}$. 

The reason why we choose the Euler-Dirac operator instead of the classical
Dirac operator is that the former has better functorial properties. In
particular, it is invariant under inversions of loops. If we consider for
example the Abelian case, $G=S^1$, and parameterize $S^1$ by $\theta\in [0,2\p]$,
then the Dirac operator reads
\ba
\displaystyle{\not} \cd = {\rm{i}}\frac{\pa}{\pa \theta}\;,
\label{D}
\ea
which, under inversion of the underlying loop 
\ba
G\to G^{-1}
\label{inve}
\ea 
picks up a minus sign. On the
other hand, we have just argued that the Euler-Dirac operator is invariant
under inversions. 

It is of course desirable to work
out a construction that works for the classical Dirac operator, but for now we
choose to work with the easier Euler-Dirac operator.\\

The particular choice of ``Dirac'' operator in (\ref{D}) is motivated by its resemblance
to a (integrated) functional derivation. Heuristically: A (smooth) connection is determined by holonomies along hoops. In the projective
system described here we consider first a finite number of hoops and a
connection is thus described 'coarse-grained' by assigning group elements to each of the
finitely many elementary
hoops. The Euler-Dirac operator (\ref{EDi}) takes the derivative on each of these copies of the
group $G$ and throws it into the Clifford bundle. In this way the Dirac
operator resembles a functional derivation operator.  

We interpret this Euler-Dirac operator as intrinsically 'quantum'
since it bears some resemblance to a canonical conjugate of the
connection. Heuristically, we write
\ba
\displaystyle{\not} \cd \sim\frac{\d}{\d\nabla}
\label{heu1}
\ea
and 
\ba
H_L \sim 1 + \nabla
\label{heu2}
\ea
due to $H_L$'s relation to the holonomy map. Here $\nabla$ is a
connection. From (\ref{heu1}) and (\ref{heu2}) the non-vanishing commutator
\[
[\displaystyle{\not} \cd,H_L] \not = 0
\]
obtains, on a very heuristical level, a resemblance to a commutation relation
of canonical conjugate variables. Thus, it is not the noncommutativity of the algebra of
holonomy loops
(to be defined rigorously below) which is 'quantum' but rather the Dirac
operator and its interaction with the algebra. This is an essential point for the interpretation of the geometrical
construction presented.

\subsection{The algebra}

We will construct our algebra as an algebra of operators on $\ch'_{si}$, or
rather a variant of hereof denoted $\ch_{si}$. This algebra will be similar, but not equal,
to the group algebra $C^*(\cl_{x^o})$ of hoops. 

 The hilbert space $\ch_{si}$ is constructed the same way as $\ch'_{si}$ but
 where $$L^2(G^{n(F)},Cl(TG^{n(F)})\otimes M_n(\bbC))$$  is used instead of
 $L^2(G^{n(F)},Cl(TG^{n(F)}))$. Here $n$ is the size of the representation of $G$. The reason for the additional matrix factor is
 that we wish to represent the holonomy loops by left matrix
 multiplication. 

 The decomposition analogous to (\ref{h2}) looks like
$$\ch_{si}=M_n (\bbC) \oplus (\oplus_{l^1}\ch_2 \otimes M_n(\bbC))\oplus
 (\oplus_{l^2} \ch_2 \otimes \ch_2 \otimes M_n(\bbC ))\ldots  \;.$$

If we are given a simple hoop $L$, we construct an operator $H_L$ on
$\ch_{si}$ in the following way:  For  a subgroup $F\in \cf_I$ we make use of
the identification (\ref{identification}) of $G^{n(F)}$ with $\Hom (F,G)$ and hence define
$$H_L (s)(\varphi )= (id \otimes \varphi(L) )(s(\varphi ))\;,$$
where $s\in  L^2(G^{n(F)},Cl(TG^{n(F)})\otimes M_n(\bbC))$ and where $\varphi
(L)=id$ when $L\notin F$. Since $H_L$ respects the maps $P_{F_1,F_2}^*$ we get
an operator $H_L$ on $\ch_{si}$. 

For a general hoop $L$, using the unique decomposition of $L$ into simple hoops $L_1\circ \ldots \circ L_n$  define $$H_L=H_{L_1}\circ \cdots \circ H_{L_n}\;.$$ 
Our algebra, which we denote $A$, is the $C^*$-algebra generated by the operators $H_L$, $L\in
\ch\cg$.  \\

It is important to realize that the
algebra $A$ is not identical to the $C^*$-algebra $C^*(\cl_{x^o})$
introduced in section \ref{Rep}. That is, we have not obtained a
representation of the group algebra of hoops on $M$. To illustrate this
consider the following two
situations:
\begin{enumerate}
\item {\it Loops with common line segment.} We consider for example two
  loops $L_1$ and $L_2$ where 
\[
L_1 = P_1\circ P_2\;,\qquad L_2 = P_2^*\circ P_3\;,
\]
with $P_i\in\cp$. Hence
\[
L_3\equiv L_1\circ L_2 = P_1\circ P_3\;.
\] 
\item {\it Intersecting loops.} Consider two loops $L_4$ and $L_5$
  where
\[
L_4(t_1)=L_5(t_2)\not= x^o\;.
\]
\end{enumerate}
In the first case $L_1$, $L_2$ and $L_3$ cannot belong to the same subgroup
$F\in\cf_I$ since they all have common line segments. Thus, their associated operators $H_{L_i}$ act on
different parts of the hilbert space. This means that they commute
\[
H_{L_1}\cdot H_{L_2}= H_{L_2}\cdot H_{L_1}\;.
\] 
In particular, it means that
\[
H_{L_1}\cdot H_{L_2}\not= H_{L_3}\;.
\]
In the second case, the product $L_4\circ L_5$ does not even belong to any
subgroup $F\in\cf_I$. Thus, the operator $H_{L_4\circ L_5}$ only exist as the
composition $H_{L_4}\cdot H_{L_5}$.

\subsection{An extended Euler-Dirac operator}

The Euler-Dirac operator defined in equation (\ref{EDi}) acts, basically, on the hilbert
space $\ch'_{si}$. When acting on $\ch_{si}$ it does not 'see' the matrix
part of the hilbert space. This need not be so. We can for example define an extended
Euler-Dirac operator by
\ba
\displaystyle{\not} \cd_{ext}(\xi \otimes m) (g) = \displaystyle{\not} \cd
(\xi (g))\otimes m + \xi (g)\otimes m_n(g)\cdot m\;,
\label{EU2}
\ea
where $m_n(g)$ is a matrix valued function on $G^{n(F)}$ and $\xi\otimes m \in
\ch_{si}$. The form of the operator in equation (\ref{EU2}) is similar to the Dirac operators of the almost commutative
geometries (including the standard model). See for example \cite{Connes:1996gi}.

\section{The space of connections}
\label{section-connection}

So far, we have considered a geometrical structure over spaces related to
certain loop group homomorphisms. We now want to describe in more detail the role of
connections in this construction.

In the above we constructed the hilbert space 
\[
\ch_{si} = \lim_{\rightarrow} L^2 (G^{m},Cl(TG^{m})\otimes M_{n}(\bbC))\;.
\]
Let us for simplicity now consider the same hilbert space but without the spin
structure and the matrix factor:
\[
\ch = \lim_{\rightarrow} L^2(G^n)\;.
\]
Hence, $\ch_{si}$ is $\ch$ with coefficients in an infinite dimensional
Clifford algebra tensored with $n$ by $n$ matrices. If we backtrack our line of reasoning we
first make the identification
\[
\ch = L^2(\lim_{\leftarrow} Hom(F,G))\;,
\]
where $F\in\cf_I$. Let $\nabla$ be a fixed, smooth connection in $\ca$. As already
mentioned, for a
given $F$, $\nabla$ gives rise to a homomorphism into $G$ via the holonomy
loop 
\[
\nabla: L \rightarrow Hol(L,\nabla)\in G\;,
\]
where $L\in F$. It is easy to see that this commutes with the structure map
and hence that we get a map
\[
\ca \to \lim_{\leftarrow} Hom(F,G)\;.
\]
Clearly, this map is injective. We therefore conclude that $\ch$
is a hilbert space over a space which contains all smooth connections.

\subsection{Distances on $\bar{\ca}$}

On a Riemannian spin-geometry the Dirac operator $D$ contains the geometrical
information of the manifold $\cm$. In particular, distances can be formulated in a
purely algebraic fashion due to Connes \cite{ConnesBook}. Given two points $x,y\in\cm$ their
distance is given by
\ba
d(x,y)=\sup_{f\in C^\infty(\cm)}\{ |f(x)-f(y)| \| [D,f]\|\leq 1    \}
\label{distance}
\ea
On a noncommutative geometry the state space replaces the notion of points. It
is possible to extend the notion of distance to the state space by
generalizing (\ref{distance}) in an obvious manner. 

For the present case, however, it is quite unclear in what sense a Dirac
operator incorporates a distance. Further, the usefulness of such a notion is
in the present situation
not obvious. Clearly, if the Dirac operator (\ref{EDi}) is interpreted as a
metric it will give rise to distances on the space $\bar{\ca}$. 

For example, it is not difficult to see that for the $G =U(1)$ case the
distance between two smooth connections will be infinite. This can be seen by first
noting that the distance induced by the Dirac operator on $U(1)^n$ is just the
sum of distances on each copy of $U(1)$. This product distance of two smooth connections will differ on infinitely many
non-intersecting loops. Further, summing these differences will give an
infinite distance between the points. Perhaps this is not
so surprising considering the fact that our geometry is infinite dimensional.

\section{Diffeomorphism invariance}
\label{section-diffeomorphism}

Clearly, the construction considered so far is very large. In fact, the
hilbert space $\ch_{si}$ is not separable and it is unclear how to extract
physical quantities in a well-defined manner. What is missing is of course
the implementation of 
diffeomorphism invariance relative to the underlying manifold
$M$. Invariance under arbitrary coordinate transformations is the defining
symmetry of general relativity and it is therefore an essential ingredients in
the formalism. It turns
out that the 'size' of the construction can indeed be drastically reduced by
taking diffeomorphism invariance into account\footnote{The construction on
  this section works both for diffeomorphisms and for extended
  diffeomorphisms, and we will therefore notationally not distinguish between
  them. But only the latter case give a seperable hilbert space, and hence the
verification of (or lack of) the axioms of a spectral triple only makes sense
for the extended diffeomorphisms.}. First we write down transformation laws of hilbert
states and operators. Next, we define diffeomorphism invariant states via a
formal sum over states connected via diffeomorphisms. We are able to represent
loop operators on such `smeared' states albeit not as a
representation of the operator algebra $A$. In a subsequent subsection we
investigate an alternative approach where we introduce an equivalence of spectral
  triples to cut down the size of both the hilbert space and the algebra as
well as the corresponding Euler-Dirac operator simultaneously. We find that
the two approaches are in fact equivalent. Finally we look at the
spectrum of the relevant Euler-Dirac operators and show that it is not fully a Dirac
operator in the sense of Connes.

We assume that the space-time dimension of the manifold $M$ is larger than
three. Since there exist no knot theory outside 3 dimensions we hereby avoid
considering different ``knot states'' etc.

\subsection{Transformations of states and operators}

We first consider states in $L^2(G^{n(F)},Cl(TG^{n(F)})\otimes M_n(\bbC))$ which are polynomial
in $g_1,\ldots,g_{n(F)}$ tensored with constant elements in $Cl(TG^{n(F)})$. A diffeomorphism $d\in Diff(M)$ which maps
\[
d:L_i\to L_i'
\]
has a natural action on such polynomials
\ba
d: p(g_1,\ldots,g_{n(F)})\to p(g_1',\ldots,g'_{n(F)})\;,
\label{d}
\ea
where $g_i'\in G'_i$ is the group corresponding to the new loop $L_i'$. 
Because we interpret states in $\ch_{si}$ as (polynomials in) holonomy loops we
can really only state how polynomials and their closure should transform under
diffeomorphisms. However, we can simply extend the transformation law
(\ref{d}) to all of $L^2(G^{n(F)},Cl(TG^{n(F)})\otimes M_n(\bbC))$ 
\[
d: \xi(g_1,\ldots,g_{n(F)}) \to \xi(g'_1,\ldots,g'_{n(F)})\;,
\]
and via the inductive limit to all of $\ch_{si}$. 

The action of the diffeomorphism group on the algebra $A$ is straight forward,
simply taken from (\ref{d}).

Above and in the following we only consider diffeomorphisms in $Diff(M)$ which
preserve the basepoint $x^{0}$.

\subsection{Diffeomorphism invariant states}
\label{sec-diff1}

From one point of view we need to solve the diffeomorphism constraint
\ba
d \xi = \xi\;,\quad \forall\; d\in Diff(M)\;;\;\;\xi\in\ch_{si}\;.
\label{diffconstraint}
\ea
Let us start by investigating this. The following is inspired by \cite{Ashtekar:2004eh,Ashtekar:1995zh}. Equation
(\ref{diffconstraint}) has, of course, the formal solution 
\ba \label{formal}
\tilde{\xi}=\sum_{d\in Diff(M)} d(\xi)\;.
\ea
This, however, makes no sense in $\ch_{si}$. Instead we need to consider the
dual of $\ch_{si}$. So, given a vector $\eta \in \ch_{si}$ we let the formal sum
(\ref{formal}) act on $\eta$ like
\ba \label{defdif}
\tilde{\xi} ( \eta)=\sum_{d \in Diff(M)} \langle d(\xi)| \eta \rangle\;.
\ea
Strictly speaking this does not make sense either, since the sum on the right
hand side need not be convergent. If we however define the action of
$\tilde{\xi}$ only on the algebraic part of $\ch_{si}$, i.e. only finite sums
of elements in the sum (\ref{h2}), the sum (\ref{defdif}) becomes finite if
the summation over $Diff (M)$ is understood correctly. We will now describe how this works: 

First we define the projection onto symmetrized states. Given a state
$\xi\in\ch_2^{\otimes n(F)}\otimes M_n$ we denote by $Diff(M|F)$
diffeomorphisms which preserve form as well as orientation of all loops in
$F$. Consider next diffeomorphisms $F\to F$ which do not lie in
$Diff(M|F)$. We denote these by $Diff(F\to F)$. The symmetry group of $F$, denoted
$SG_F$, is the quotient 
\ba
SG_F = Diff(F\to F)/Diff(M|F)\;.
\label{SG}
\ea
They consist of certain permutations and inversions. The projection is
defined by
\ba
P(\xi) = \frac{1}{N_F}\sum_{d \in SG_F} d( \xi )\;,
\label{projection}
\ea
where $N_F$ is the number of elements in $SG_F$. Next, consider the remaining
diffeomorphisms which moves the loops in $F$ outside $F$. We define the sum
(\ref{defdif}) by
\ba
\label{sum}
\tilde{\xi} ( \eta)=\sum_{d \in Diff(M)/Diff(F\to F)} \langle d ( P\xi)| \eta \rangle\;,
\ea
where the sum is interpreted as an effective sum, i.e. if $d_1(F)=d_2(F)$ we
identify $d_1$ and $d_2$. 
If $\eta\in \ch_{2}^{\otimes n(F)}\otimes M_n$ we find $N_F$ contributions on
the rhs of
(\ref{sum}). Else it is zero.

The vector space of linear combinations of sums (\ref{sum}) is given the inner
product
\ba
\langle \tilde{\xi}_1| \tilde{\xi}_2\rangle = \tilde{\xi}_1(\xi_2)\;.
\label{norm}
\ea
The crucial point is that this sum has finitely many non-vanishing terms (see above). 

The completion of this vector space in the norm (\ref{norm}) is a
diffeomorphism invariant hilbert space which we denote by $\ch_{diff}$.

The problem with this construction is that it is somewhat unclear how the algebra
of hoops should be represented on $\ch_{diff}$. Since our goal is to find a spectral triple involving not
only a separable hilbert space but also a (separable) algebra and a well defined
Dirac operator, this is clearly a crucial point. The difficulty stems from the
fact that the algebra is not diffeomorphism invariant but rather
co-variant. The Dirac operator, on the other hand, is diffeomorphism
invariant and therefore causes no problems. 

Essentially, we need to make sense of a 'smearing' of algebra
elements according to
\ba
\tilde{H}_L = \sum_{d\in Diff(M)} H_{d(L)}\;,
\label{stands}
\ea
similar to equation (\ref{formal}). As it stands, equation (\ref{stands}) is
meaningless. Instead we do the following: Given a hoop operator $H_L\in A$
define the symmetrized operator by
\ba
P_F(H_L) = \frac{1}{N_F}\sum_{d \in SG_F}H_{d(L)} \;,
\label{smeared-operator}
\ea
where $SG_F$ is the symmetry group of a subgroup $F$ including $L$. $N_F$ is again the total
number of elements in $SG_F$. For example, if $L$ is simple and $F$ is the
algebra generated by $L$, we have 
\[
P_F(H_L)= \frac{1}{2}\left(L + L^{-1}\right)\;.
\]
For a 'smeared' state $\tilde{\xi}\in \ch_{diff}$  we define the action of
$H_L$ on $\tilde{\xi}$ by
\[
H_L (\tilde{\xi}) = \sum_{d \in Diff(M)/Diff(F\to F)} d(P_F(H_L)\cdot  P(\xi))\;,
\]
where we choose the representative $\xi$ so that $L$ and $\xi$ have coinciding
domains and where $P_F$ is taken with respect to the subgroup $F$ defined by
the domain of $\xi$ and $L$.

Note that we no longer deal with a representation of loops. For example, given
a simple loop $L$ acting on a state $\xi$ with domain on a single copy of $G$
we find that (using a somewhat sloppy notation)
\[
H_L \cdot H_L = \tfrac{1}{4}(H_{L^2} + H_{L^{-2}} + 2)\;.
\]
This relation, however, changes according to what states in $\ch_{diff}$ $H_L$ acts on.

\subsection{Diffeomorphism invariance via equivalent triples}

In the previous subsection we implemented diffeomorphism invariance by
constructing diffeomorphism invariant states and defining an action of loop
operators hereon. In fact, there is another option which, however, only works
for extended diffeomorphisms. As explained above, the diffeomorphism group
acts not only on the hilbert space but also on the algebra. We can therefore
define an equivalence on the level of sub-triples; algebra, hilbert space and
Euler-Dirac operator. This identification happens at the level of subgroups $F\in\ch\cg$.

If we consider a single, simple loop $L$, the spectral triple associated to
this is just 
\ba
(<L>, L^2(G,Cl(TG)\otimes M_n),\displaystyle{\not} \cd)\;,
\label{st1}
\ea
where $<a,b,\ldots>$ is the $C^*$-algebra generated by $\{a,b,\ldots\}$. Since
all single, simple loops are diffeomorphic, at this level we just get expression
(\ref{st1}) when we identify spectral sub-triples which are diffeomorphic. At
the level of two nonintersecting simple loops $L_1$ and $L_2$ the spectral
triple associated to this is
\ba
(<L_1,L_2>, L^2(G^2,Cl(TG^2)\otimes M_n),\displaystyle{\not} \cd)\;.
\label{st2}
\ea
Again, by identifying spectral triples of diffeomorphic loops we get at this
level just expression (\ref{st2}). This picture simply continues for all finitely
generated subgroups and taking the limit hereof gives us an equivalence class
of spectral triples represented by the infinite dimensional triple
\ba
(<L_1,L_2,\ldots>, L^2(G^\infty,Cl(TG^\infty)\otimes M_n,\displaystyle{\not} \cd)\;.
\label{triple}
\ea

Further, not only are all subgroups of nonintersecting loops with $n$
generators diffeomorphic. There are also internal diffeomorphisms which
shuffle the generators. One can factor out this symmetry by symmetrizing
operators and states, just as we did in the previous subsection.

Therefore, the result is, in fact, identical to the result of the previous subsection.

Instead of symmetrizing one could also make the noncommutative quotient of the
action of the internal diffeomorphism group $SG_F$ (and the limit),
i.e. consider the crossed product $A_F\times SG_F$, where $A_F$ is the part of
our algebra acting on the $F$ part. This would be more in the spirit of
noncommutative geometry and Connes. We will investigate this alternative elsewhere.

\subsection{Spectral in the sense of Connes?}

It remains to clarify whether the spectral triple (\ref{triple}) satisfy the
conditions put forward by Connes \cite{ConnesBook}, see also \cite{Moscovici}. A confirmative
answer will permit us the full power of noncommutative geometry. 
 
Clearly, on each level in the projective/inductive limit, the relevant Dirac
(Euler-Dirac) operator satisfy the conditions for a spectral triple, simply
per construction. The question remains whether it also holds in the limit.

There are three conditions. 
First, the operator
\ba
[\displaystyle{\not} \cd,a]\;,
\label{comm}
\ea
where $a$ belongs to the subspace of $A$ of finite linear combinations of loop
operators, has to be bounded. 
A simple loop operator $a = H_L$, acts, according
to (\ref{smeared-operator}), on a state via 
\ba
\frac{1}{N_F} (H_{L_1}+ H_{L^{-1}_1}+\ldots +H_{L_{n_F}}+H_{L^{-1}_{n(F)}})\;,
\label{1}
\ea
where the number $n(F)$ refers to the domain of $L$ and the state on which it
acts (see section \ref{sec-diff1}). We can estimate the commutator
of (\ref{1}) with the Dirac operator by
\ba
&&\frac{1}{N_F} \|\big[\displaystyle{\not} \cd,H_{L_1}+ H_{L^{-1}_1}+\ldots
  +H_{L_{n(F)}}+H_{L^{-1}_{n(F)}}\big]
\|\nonumber\\ 
&\leq& 
\frac{1}{N_F} \big(
\|\big[\displaystyle{\not} \cd,H_{L_1}\big]\| +
\|\big[\displaystyle{\not} \cd, H_{L^{-1}_1}\big]\|+\ldots
  +\|\big[\displaystyle{\not} \cd,H_{L_{n_F}}\big]\|+\|\big[\displaystyle{\not}
    \cd,H_{L^{-1}_{n(F)}}\big]\|\big)\nonumber\\ 
&=& \|\big[\displaystyle{\not} \cd,H_{L}\big]\|\;.
\ea
Because the operator 
\[
H_L: G\to G\;;\quad g\to g\;
\]
is bounded we conclude that the operator (\ref{comm}) is bounded for a simple loop
operator. For compositions of simple loop operators the arguments is repeated
and therefore we conclude that the first condition is satisfied.

Second, we need to investigate
whether the operator 
\[
\frac{1}{\displaystyle{\not} \cd-\l}\;,\quad \l\in\bbC/\bbR
\]
is compact. In fact, this turns out not to be the case. Let us explain. 
For simplicity we leave out the matrix part of the hilbert space and
simply consider the space
\[
L^2(G^n,Cl(TG^n)) = L^2(G,Cl(TG))^{\otimes n}\;,
\]
where we only consider symmetrized (un-ordered) elements according to (\ref{projection}). Given a set of
eigenfunctions $\{\xi_1,\ldots,\xi_m \}$ in $L^2(G,Cl(TG))$ of the Dirac
operator, the product
\ba
\xi_{i_1}\otimes\ldots\otimes\xi_{i_n}
\label{eigenstates}
\ea
is an eigenfunction of the Dirac operator in $L^2(G^n,Cl(TG^n))$. The problem
is that if we find a function $\xi_0$ in $L^2(G,Cl(TG))$ with eigenvalue zero
and which differs from the function $1$,
then we will automatically have an infinite dimensional eigenspace associated
to any eigenvalue. To see this simply consider the function
(\ref{eigenstates}) (remember that we consider only symmetrized products)
\[
\xi_0\otimes\xi_{i_1}\otimes\ldots\otimes\xi_{i_n}
\]
in $L^2(G^{(n+1)},Cl(TG^{(n+1)}))$. This is again an eigenfunction with the same eigenvalue as
(\ref{eigenstates}). 

According to Hodge theory (see theorem II.5.15 in \cite{spingeometri}) the
kernel of a Euler-Dirac operator on a compact manifold $M$ is related to the
cohomology group:
\[
ker(\displaystyle{\not} \cd) = \oplus \bbH^p\;,\quad \bbH^p = H^p(M;\bbC)\;,
\]
and the cohomology group is, at least on an orientable manifold as the
Lie group $G$, not empty (the volume form is an example). Therefore we
conclude that the Euler-Dirac operator in (\ref{triple}) does not satisfy Connes'
second condition.

In principle, it is possible to correct this ``flaw'' in the construction of the Dirac
operator in (\ref{triple}) by adding a bounded perturbation to
$\displaystyle{\not} \cd$ on each level in the projective/inductive
limit. Such a perturbation will, in general, not be bounded in the limit
itself. Indeed, if the perturbation is constructed in a way so that the
perturbed Dirac operator satisfy condition two, then the full perturbation
will be unbounded.

Changing the operator on each level of the projective/inductive limit does not
change the K-homology class at each level. In the limit, however, the operator
will be changed
(the original Euler-Dirac operator in (\ref{triple}) does not have a
K-homology class). 

The third condition is self-adjointness. That $\displaystyle{\not} \cd$ is
self-adjoint is secured by construction.

Let us end this subsection by noting that the fact that the Dirac operator in the triple (\ref{triple}) does not satisfy
Connes second condition may be interpreted as a hint that there exist some
extra symmetries that have not been (and should be) factored out.

\section{Discussion \& outlook}
\label{sec-dis}

In the present paper we presented new ideas on the unification of
noncommutative geometry -- in particularly Connes formulation of the standard
model -- and the principles of quantum field theory. We apply the machinery
of noncommutative geometry to a general function space of connections related to
gravity. A noncommutative algebra of holonomy loops is represented on a
separable, diffeomorphism invariant hilbert space. An Euler-Dirac operator is
constructed. The whole setup relies on techniques of projective and inductive
limits of algebras, hilbert spaces and operators.\\

What comes out is a geometrical structure, including integration theory, on a
space of field configurations modulo diffeomorphism invariance. A global
notion of differentiation (the Dirac operator) is obtained. We find it
remarkable that the whole construction boils down to the study of Dirac
operators on various copies of some Lie-group.  \\

Whereas the noncommutativity of the algebra is intrinsically classic we
interpret the Dirac operator, which resembles a functional derivation, as
'quantum'. \\

Certain problems arose during the analysis. First, we were unable to represent
the full hoop group in a manner compatible with the Euler-Dirac operator. The
solution proposed and analyzed is to consider only finite subgroups of
non-intersecting loops (in the projective system). This modification has
important consequences; instead of graphs (spin-networks) we deal with
polynomials on various copies of the group. It is, however, not clear to us
whether this is an important point.

More seriously, the final Dirac operator does not fulfill the conditions
formulated by 
Connes. In particular, it has infinite-dimensional eigenspaces. 
Thus, we did not succeed to construct a spectral triple which satisfy the
conditions put forward by Connes.\\

A prime concern to further development is to understand why our
constructed Dirac operator is not spectral in the sense of Connes. We suspect
that what is missing is a symmetry related to the infinite dimensional
Clifford algebra, i.e. $Cl(T_{id}(G^\infty))$. Another possible solution is to
use the ordinary Dirac operator instead of the Euler-Dirac operator. To
account for lack of invariance under inversions of loops one can double the
hilbert space: Instead of for each simple loop to assign the hilbert space of
square integrable functions over $G$ we can assign two copies of this hilbert
space; one for each orientation of the loop. The diffeomorphism associated
with inversion of the loop will then act by interchanging the two hilbert
spaces. There will however still be some problems, for example embedding
properties when we increase the the number of copies of $G$'es.

Another concern is to extend the present construction to work for non-compact
groups, since gravity involves $SO(3,1)$. The main problem will be the embeddings in the projective limit. For example, $L^2(G)$
is not naturally embedded in in $L^2(G^2)$ whenever $G$ is non-compact. We believe, however, that this is
a technical and solvable problem. Also, loops will no longer occur as
states in the hilbert space; a priori not necessarily a problem. \\

Also, we would like to understand in what sense the noncommutativity of the
holonomy algebra generates a bosonic sector and, if so, what it is. Clearly,
noncommutativity permits inner automorphisms and nontrivial fluctuations of
the Dirac operator. If we assume that we succeed to construct a Dirac operator $D$
satisfying all of Connes' conditions, and if we consider fluctuations around $D$ of the form
\ba
D\to \tilde{D} = D + A + J A J^\dagger\;,
\nonumber
\ea
where J is Tomita's anti-linear isometry \cite{Ta} and $A$ is a noncommutative
one-form\footnote{elements of $\Omega^n_D$ are of the form $a_0
  [D,a_1]\ldots [D,a_n]$ where the $a_i$'s are elements of the algebra \cite{Connes:1996gi}.}, $A\in\Omega^1_D$,
 then we can apply Chamseddine and Connes'
spectral action principle \cite{Chamseddine:1991qh,Chamseddine:1996rw}. Thus,
we can write down automorphism invariant quantities like
\ba
\langle \tilde{\xi}| \tilde{D} | \tilde{\xi}\rangle\;,\quad
\mbox{Tr} \varphi \big(\tilde{D}\big) \;,\quad\ldots \;.
\nonumber
\ea
Such terms can be interpreted as integrated quantities, schematically, of the form
\[
\int_{\bar{\ca}/Diff}d\nabla \ldots
\]
which resembles a Feynman path integral and
contains both fermionic and bosonic degrees of freedom. Here the
integration is defined, modulo diffeomorphisms, on a space of connections.  \\

In the introduction we motivated our analysis by stating that Connes
formulation of the standard model coupled to gravity is intrinsically classical. With the aim of
combining noncommutative geometry and the principles of quantum field theory,
we have found a spectral triple which a priori appears to be quite far from field
theory. It is clearly of prime concern to investigate whether the construction
does contain a field theory limit and, if so, what it is.

\section{Acknowledgment}

It is a pleasure to thank Raimar Wulkenhaar for comments and for carefully reading the manuscript.

\appendix

\section{Clifford algebras and Dirac operators}
\label{diracappendix}

Here we give a brief review of Clifford algebras and the Euler-Dirac
operator. For a detailed account see for example \cite{spingeometri}.
Since we are interested in Clifford algebras over Lie-groups we only treat the
Euclidian case.

Let $V$ be a real vector-space. We define the tensor-algebra $T(V)$ as
$$T(V)=\sum_{i\geq 0} V^{\otimes^i}$$
with multiplication 
$$v_1\otimes \ldots \otimes v_n \cdot u_1\otimes \ldots \otimes u_m =v_1\otimes \ldots \otimes v_n \otimes u_1 \otimes \ldots \otimes u_m\;.$$

Given a metric $\langle\cdot , \cdot \rangle$ on $V$ one defines the Clifford algebra as
$$Cl(V)=T(V)/(v\otimes u+u\otimes v =-2\langle v,u\rangle )\;.$$ 

If $e_1,\ldots , e_n$ is an orthonormal basis of $V$ the Clifford algebra $Cl(V)$ consists of elements on the form  
$$e_{i_1}\cdots e_{i_k}\;,$$
where $i_1< \cdots <i_k$ and with the product rules
$$e_ie_j=-e_je_i, \quad i\not= j , \quad e_i^2=-1\;.$$ 

There is an inner product on $Cl(V)$ given by
$$\langle e_{i_1}\cdots e_{i_k} , e_{j_1}\cdots e_{j_l}\rangle =1$$
if $k=l$ and $i_1=j_1,\ldots ,i_k=j_l$ and zero else.

The group $O(n)$ acts on $Cl(V)$ by 
$$o(e_{i_1}\cdots e_{i_k})=o(e_{i_1})\cdots o(e_{i_k}),\quad o\in O(n)$$
as automorphisms preserving the inner product. In particular one also gets an action of $\frak{so}(n)$ on $Cl(V)$.

For a manifold $M$ with a metric, one defines the Clifford bundle $Cl(TM)$ as 
the bundle 
$$M\ni m \to Cl(T_mM )\;,$$
where the inner product on $T_mM$ is the one given by the metric. 

Let $\nabla$ denote the Levi-Civita connection associated to the metric. Via the extension of the action of $O(n)$ from $V$ to $Cl(V)$, the Levi-Civita connection extends to a connection in $Cl(TM)$ via the formula
$$\nabla (e_{i_1}\cdots e_{i_k})=\sum_l e_{i_1}\cdots \nabla (e_{i_l})\cdots e_{i_k}\;,$$
where $e_{i_l}$ are local orthonormal sections in $TM$.

One defines the Euler Dirac operator $D$ by
$$L^2(M,Cl(TM))\ni s \to D(s)=\sum_i e_i\cdot \nabla_{e_i}s\,,$$
where $\{ e_i \}$ is a local orthonormal sections in $TM$.   
  
Since one wants to work with hilbert spaces one complexifies  the space $L^2(M,Cl(TM))$ leaving the notion unchanged.

\section{Projective and inductive limits}
\label{proj}

Here we review the concepts of projective and inductive
limits. For a different treatment we refer to \cite{Marolf:1994cj}.

\subsection{Projective limits}
To illustrate the concept of a projective limit we will consider the index set $\bbN$ and for each $n\in \bbN$ the space $\bbR^n$. If $n_1\leq n_2$ there are projection
$$P_{n_2,n_1}: \bbR^{n_2}\to \bbR^{n_1}$$ 
given by
$$P_{n_2,n_1} (x_1,\ldots,x_{n_2})=(x_1,\ldots , x_{n_1})\;.$$
We define the product 
$$\prod_{n\in \bbN}\bbR^n=\{ (X_n)_{n\in \bbN} | X_n\in \bbR^n\}\;,$$
i.e. an element in  $\prod_{n\in  \bbN}\bbR^n$ is just where we pick an element in each $\bbR^n$ for all $n$. An element can thus be written as 
$$(x_1^1,(x_1^2,x_2^2),(x_1^3,x_2^3,x_3^3),\ldots)\;.$$ 

The projective limit is defined as those elements in $\prod_{n\in \bbN}\bbR^n$ where
$$(x_1^{n_1},\ldots , x_{n_1}^{n_1})= P_{n_2,n_1}(x_1^{n_2},\ldots , x_{n_2}^{n_2})\;.$$
or written out
$$x_1^1=x_1^2, \quad (x_1^2,x_2^2)=(x_1^3,x_2^3 ), \quad (x_1^3,x_2^3,x_3^3)=(x_1^4,x_2^4,x_3^4),\ldots$$
In other words, the projective limit, also written 
$$\lim_\leftarrow (\bbR^n , P_{n_2,n_1})\;,$$
is just $\bbR^\infty$, the set of all sequences in $\bbR$.

Another example which is more relevant to our case, comes from group
theory. Let $G$ be a group. We let $\cf$ be the set of finitely generated subgroups of $G$. If $F_1,F_2 \in \cf$ and $F_1\subset F_2$ we have the inclusion map $\iota_{F_1,F_2}: F_1\to F_2$. If we therefore consider group homomorphism from each of these finitely generated subgroups to a fixed group $G_1$ we get, by dualizing, restriction maps
$$\iota_{F_1,F_2}^*: Hom (F_2,G_1)\rightarrow Hom (F_1,G_1)\;.$$ 
As in the case of $\bbR^n$ we can consider the product 
$$\prod_{F\in \cf}Hom(F,G_1)=\{ (\varphi_{F})_{F\in \cf}| \varphi_{F}\in Hom(F,G_1)\}\;, $$
and the projective limit is defined as the subset of the product of sequences that are consistent with the restriction maps, i.e. a sequence $(\varphi_F)_{F\in \cf}$ is in the projective limit if 
$$\iota^*_{F_1,F_2} (\varphi_{F_2})=\varphi_{F_1}\;,$$
for all $F_1,F_2\in \cf$ with $F_1\subset F_2$.

We note that we have a  map 
$$\Phi :Hom(G,G_1)\to \lim_\leftarrow (Hom(F,G_1),\iota_{F_1,F_2}^*)$$
just by restricting a homomorphism from $G$ to $G_1$ to its finite subgroups. It is easy to see that this map is a bijection, and we can hence identify $Hom(G,G_1)$ with the projective limit. This might seem like we have just expressed something easy, namely $Hom(G,G_1)$ with something complicated, namely the projective limit. However the description as a projective limit turns out to be very useful. 

\subsection{Inductive limits}

Inductive limit is  the dual concept of projective limit. For simplicity we take $\bbT^\infty$, the infinite torus (easier than $\bbR^\infty$ since $\bbT^n$ is compact). This means that we have a projective system
$$P_{n_2,n_1} : \bbT^{n_2} \to \bbT^{n_1},\quad n_1,n_2\in \bbN, \quad n_1\leq n_2\;,$$
where $\bbT^n$ is the $n$-torus and $P_{n_2,n_1}$ are the natural projections.

The dual of a space is the functions on the space. There are of course several candidates for functions. In this example we will take the space of square integrable functions on $\bbT^n$ with respect to the Haar measure, i.e. $L^2 (\bbT^n,d\mu_H)$. The dual map of $P_{n_2,n_1}$ gives a map
$$P_{n_2,n_1}^*:L^2(\bbT^{n_1})\to L^2(\bbT^{n_2})$$ 
 defined by
$$P_{n_2,n_1}^* (\xi )(x)=\xi(P_{n_2,n_1}(x )), \quad x\in \bbT^{n_2}\;.$$
These maps are embeddings and are maps of hilbert spaces since 
$$\int 1 d\mu_H=1\;.$$
The inductive limit of these hilbert spaces are constructed in the following way: We take the direct sum 
$$\oplus_n L^2(\bbT^n)\;,$$
i.e. sequences $\{\xi_n\}_{n\in \bbN}$ with $\xi_n\in L^2(\bbT^n)$ such that 
$\{\xi_n\}$ is zero from a certain step. In this space we consider the subspace $N$ generated by elements of the form
$$(0,\ldots ,0, \xi_{n_1},0,\ldots,0,-P_{n_2,n_1}^* (\xi_{n_1}) ,0,\ldots)\;,$$
and form the quotient space $\oplus_n L^2(\bbT^n)/N$. This quotient just means that we consider all vectors lying in some $L^2(\bbT^n)$, and identify two vectors $\xi_{n_1}, \xi_{n_2}$ if $P_{n_2,n_1}^*(\xi_{n_1})=\xi_{n_2}$. The space 
$$\oplus_n L^2(\bbT^n)/N$$
is the algebraic inductive limit 
$$\lim_{\to}L^2(\bbT^n)\;.$$

Naively we are considering $L^2(\bbT^1)$ as a subspace of $L^2(\bbT^2)$, $L^2(\bbT^2)$ as a subspace of $L^2(\bbT^3)$, $L^2(\bbT^3)$ as a subspace of $L^2(\bbT^4)$ and so on, and the limit space as $n$ tends to infinity is the direct limit. Or in a picture
$$L^2(\bbT^1 )\subset L^2(\bbT^2)\subset L^2(\bbT^3 )\subset \ldots \subset
\lim_\to L^2(\bbT^n)\;.$$

We have used the word algebraic inductive limit, since we want to put some hilbert space structure on the inductive limit. 

If we have two vectors in the inductive limit, let us say $\xi_1\in L^2(\bbT^{n_1})$ and $\xi_{n_2}\in L^2(\bbT^{n_2})$ we define the inner product by:
$$<\xi_{n_1},\xi_{n_2}>=<P_{n_2,n_1}^*(\xi_{n_1}),\xi_{n_2}>_{L^2(\bbT^{n_2})}\;.$$
Since the embeddings $P_{n_2,n_1}^*$ are hilbert space maps, this inner product is well defined.

The definition of the hilbert space inductive limit of $\{L^2(\bbT^n), P_{n_2,n_1}^*\}$ is therefore the completion of $\oplus_nL^2(\bbT^n)/N$ in the inner product $<\cdot , \cdot >$. We will also denote this limit with
$$\lim_{\to}L^2(\bbT^n)\;.$$

\subsection{Constructing operators on inductive limits of hilbert spaces}
The main advantage of giving a description of spaces as projective or inductive limits is that one can work on each copy, and then extend to the hole space if the construction is compatible with the structure maps, i.e. $P_{n_2,n_1}^*$ for example. 

As an example of this, let us take \raisebox{-1.3ex}{ $\stackrel{\displaystyle{\lim}}{\to}$}$ L^2(\bbT^n)$. On $L^2(\bbT^n)$ we have the Laplacian $\Delta_n:L^2(\bbT^n)\to L^2(\bbT^n)$ defined by
$$\triangle_n=-(\partial^2_{\theta_1} +\partial^2_{\theta_2}+\ldots +\partial^2_{\theta_n})\;.$$ 

Note that 
$$P^*_{n_2,n_1}(\Delta_{n_1}(\xi_{n_1}))=\Delta_{n_2}(P^*_{n_2,n_1}(\xi_{n_1}))\;,$$
and therefore $\Delta=\sum_n\Delta_n$ on $\oplus L^2(\bbT^n)$ has the property 
$$\Delta (N)\subset N\;,$$
i.e. $\Delta$ descends to  a densely defined operator on the quotient space, i.e. the inductive limit \raisebox{-1.3ex}{ $\stackrel{\displaystyle{\lim}}{\to}$}$ L^2(\bbT^n)$.

\end{document}